
\hsize = 6in
\vsize = 8in
\hoffset=0.6cm
\voffset=1.5cm

\catcode`@=11

\font\twelverm=cmr12   
\font\ninerm=cmr9
\font\sevenrm=cmr7

\font\twelvei=cmmi12   
\font\ninei=cmmi9
\font\seveni=cmmi7

\font\twelvesy=cmsy10 scaled 1200  
\font\ninesy=cmsy9
\font\sevensy=cmsy7

\font\tenex=cmex10     

\font\twelvebf=cmbx12  
\font\ninebf=cmbx9
\font\sevenbf=cmbx7

\font\twelvett=cmtt12   

\font\twelvesl=cmsl12   

\font\twelveit=cmti12   



\font\twelvebigbf=cmbx12 scaled 1200
\font\twelveBigbf=cmbx12 scaled 1440

\skewchar\twelvei='177 \skewchar\ninei='177 \skewchar\seveni='177
\skewchar\twelvesy='60 \skewchar\ninesy='60 \skewchar\sevensy='60

\def\twelvepoint{\def\rm{\fam0\twelverm}
          \def\mit{\fam1}
          \def\oldstyle{\fam1\twelvei}
          \def\cal{\fam2}
          \def\it{\fam\itfam\twelveit}
          \def\sl{\fam\slfam\twelvesl}
          \def\bf{\fam\bffam\twelvebf}
          \def\tt{\fam\ttfam\twelvett}
          \def\bigbf{\twelvebigbf}
          \def\Bigbf{\twelveBigbf}
  \textfont0=\twelverm  \scriptfont0=\ninerm  \scriptscriptfont0=\sevenrm
  \textfont1=\twelvei   \scriptfont1=\ninei   \scriptscriptfont1=\seveni
  \textfont2=\twelvesy  \scriptfont2=\ninesy  \scriptscriptfont2=\sevensy
  \textfont3=\tenex  \scriptfont3=\tenex      \scriptscriptfont3=\tenex
  \textfont4=\twelveit
  \textfont5=\twelvesl
  \textfont6=\twelvebf  \scriptfont6=\ninebf  \scriptscriptfont6=\sevenbf
  \textfont7=\twelvett
  \parindent=30pt
  \footindentamount=\parindent
  \parskip=0pt plus 2pt
\def\normalspaces{\abovedisplayskip=15pt plus 4pt minus 4pt
                  \belowdisplayskip=15pt plus 4pt minus 4pt
                  \abovedisplayshortskip=0pt plus 4pt
                  \belowdisplayshortskip=8pt plus 3pt minus 3pt
                  \smallskipamount=3pt plus 1pt minus 1pt
                  \medskipamount=7pt plus 2pt minus 2pt
                  \bigskipamount=15pt plus 5pt minus 5pt
                  \alignskipamount=7pt plus1pt minus1pt
                  \refbetweenskipamount=8pt plus 2pt minus 1pt
                  \normalbaselineskip=17pt
                  \normallineskip=1pt
                  \parskip=0pt plus 2pt
                  \jot=3pt
                  \chaplineskipamount=0.2cm
                  \sectlineskipamount=3pt
                  \chapaboveskipamount=1.5cm plus1.5cm minus0.3cm
                  \chapbelowskipamount=0.3cm plus0.1cm
                  \sectfirstskipamount=0.1cm
                  \sectaboveskipamount=0.7cm plus 1.0cm
                  \sectbelowskipamount=0.3cm plus 0.1cm
                  \normalbaselines}
\def\doublespaces{\abovedisplayskip=22pt plus 6pt minus 6pt
                \belowdisplayskip=22pt plus 6pt minus 6pt
                \abovedisplayshortskip=0pt plus 5pt
                \belowdisplayshortskip=10pt plus 4pt minus 4pt
                \smallskipamount=5pt plus 1pt minus 1pt
                \medskipamount=10pt plus 3pt minus 3pt
                \bigskipamount=22pt plus 7pt minus 7pt
                \alignskipamount=2pt plus1pt minus1pt
                \refbetweenskipamount=14pt plus 3pt minus 1pt
                \normalbaselineskip=24pt
                \normallineskip=1pt
                \parskip=0pt plus 5pt
                \jot=4pt
                \chaplineskipamount=0.3cm
                \sectlineskipamount=5pt
                \chapaboveskipamount=2.0cm plus2.0cm
                \chapbelowskipamount=0.5cm plus0.2cm
                \sectfirstskipamount=0.15cm
                \sectaboveskipamount=1.0cm plus 1.5cm
                \sectbelowskipamount=0.5cm plus 0.2cm
                \normalbaselines}
  \setbox\strutbox=\hbox{\vrule height10.2pt depth4.2pt width0pt}
  \normalspaces\rm}

\newskip\newlineskipamount
\newskip\refbetweenskipamount
\newskip\alignskipamount

\newcount\CHAPNUM  \CHAPNUM=0
\newcount\SECTNUM  \SECTNUM=0
\newcount\REFNUM   \REFNUM=0
\newcount\EQNUM    \EQNUM=0
\newcount\APPNUM   \APPNUM=64
\newcount\EQUATIONCODE \EQUATIONCODE=0
\newcount\SHOWCODE     \SHOWCODE=0

\newskip\newlineskip
\newskip\chaplineskipamount
\newskip\sectlineskipamount
\newskip\chapaboveskipamount
\newskip\chapbelowskipamount
\newskip\sectaboveskipamount
\newskip\sectfirstskipamount
\newskip\sectbelowskipamount
\newskip\footindentamount

\newbox\REFBOX
\newbox\REFDUMPBOX

\def\vfootnote#1{\insert\footins\bgroup
  \parindent=0pt
  \interlinepenalty\interfootnotelinepenalty
  \splittopskip\ht\strutbox 
  \splitmaxdepth\dp\strutbox \floatingpenalty\@MM
  \leftskip\footindentamount \rightskip\z@skip
  \spaceskip\z@skip \xspaceskip\z@skip
  \vskip\medskipamount
  \textindent{\rm #1}\footstrut\rm\futurelet\next\fo@t}

\def\pagenumbers{\global
\footline={\hss\lower0.5cm\hbox{\twelverm\folio}\hss}}
\def\nofirstpagenumber{\global
\footline={\hss\lower0.5cm\hbox{\ifnum\pageno=1{}\else
\twelverm\folio\fi}\hss}}
\pagenumbers


\def\lapprox{\mathrel{\mathop
  {\hbox{\lower0.5ex\hbox{$\sim$}\kern-0.8em\lower-0.7ex\hbox{$<$}}}}}
\def\gapprox{\mathrel{\mathop
  {\hbox{\lower0.5ex\hbox{$\sim$}\kern-0.8em\lower-0.7ex\hbox{$>$}}}}}
\def\eg{{\it e.g.}}

\def\ie{{\it i.e.}}

\def\pn#1E#2 #3;{#1{\times}10^{#2}\,{\rm#3}}
\def\ex#1;{10^{#1}}
\def\mn#1E#2;{#1{\times}10^{#2}}

\twelvepoint
\def\alignskip{\noalign{\vskip\alignskipamount}}
\def\creturn{\par\nobreak\vskip\newlineskip
   \nobreak\hfil}

\def\title#1{{\parindent=0pt\parskip=0pt
  \newlineskip=0.3cm
  \line{\hfil}
  \vskip1cm plus2cm minus 0.2cm
  \hfil\bigbf#1\par}}
\def\author#1#2{\vskip1.0cm plus 0.3cm minus 0.3cm
  {\newlineskip=0pt\parindent=0pt\parskip=0pt
   \centerline{\rm #1}
   \vskip0.3cm
   \hfil\sl #2\par}}

\long\def\abstract#1{\vskip1.5cm plus 0.7cm minus 0.4cm
  \centerline{ABSTRACT}
  \nobreak\vskip\chapbelowskipamount\nobreak
  \noindent#1\par}
\def\newpage{\vfil\eject}

\def\chapter#1{{\parindent=0pt\parskip=0pt
  \newlineskip=\chaplineskipamount
  \vskip\chapaboveskipamount\penalty-400
  \global\advance\CHAPNUM by1 \global\SECTNUM=96
  \ifnum\EQUATIONCODE=1\global\EQNUM=0\fi
  \hfil\uppercase\expandafter{\romannumeral\the\CHAPNUM.~#1}\par}
  \nobreak\vskip\chapbelowskipamount\nobreak}
\def\section#1{{\parindent=0pt\parskip=0pt
  \global\advance\SECTNUM by1
  \ifnum\SECTNUM=97\vskip\sectfirstskipamount\nobreak\else
  \vskip\sectaboveskipamount\penalty-300\fi
  \newlineskip=\sectlineskipamount
  \hfil{\it\char\the\SECTNUM) #1}\par}
  \nobreak\vskip\sectbelowskipamount\nobreak}
\def\nochap#1{{\parindent=0pt\parskip=0pt
  \newlineskip=\chaplineskipamount
  \vskip\chapaboveskipamount\penalty-400
  \global\SECTNUM=96
  \hfil\uppercase\expandafter{#1}\par}
  \nobreak\vskip\chapbelowskipamount\nobreak}
\def\appendix#1{{\parindent=0pt\parskip=0pt
  \newlineskip=\chaplineskipamount
  \vskip\chapaboveskipamount\penalty-400
  \global\advance\APPNUM by1 \global\SECTNUM=96
  \global\EQUATIONCODE=2\global\EQNUM=0
  \hfil\uppercase\expandafter{APPENDIX~\char\the\APPNUM:\quad#1}\par}
  \nobreak\vskip\chapbelowskipamount\nobreak}

\def\eq{\global\advance\EQNUM by1\eqno(\the\EQNUM)}
\def\EQ#1{\global\advance\EQNUM by1
  \ifnum\EQUATIONCODE=0 \xdef#1{\the\EQNUM} \fi
  \ifnum\EQUATIONCODE=1 \xdef#1{\the\CHAPNUM.\the\EQNUM} \fi
  \ifnum\EQUATIONCODE=2 \xdef#1{{\rm\char\the\APPNUM.\the\EQNUM}}\fi
  \ifnum\SHOWCODE=0 \eqno(#1) \fi
  \ifnum\SHOWCODE=1 \eqno(#1)\hbox to0pt{\tt\ \string#1\hss}\fi}
\def\frac#1#2{{#1\over#2}}
\def\half{{\textstyle{1\over2}}}

\catcode`@=12 

\setbox\REFBOX=\vbox{\nochap{References}}
\def\RF#1#2{\gdef#1{\global\advance\REFNUM by1
  \global\setbox\REFBOX=\vbox{\unvbox\REFBOX
  \vskip\refbetweenskipamount\goodbreak
  {\leftskip=0pt\rightskip=0pt\rm\item{\the\REFNUM.}#2\par}}
  \the\REFNUM\xdef#1{\the\REFNUM}}}
\def\qref#1{\unskip$^{#1}$}
\def\refout{\vskip1.5cm plus2cm\penalty-400
  \unvbox\REFBOX}
\def\dumpref#1{\setbox\REFDUMPBOX=\vbox{#1}-}
\def\journal#1; #2;{\gdef#1##1 (##2) ##3;{{\sl#2\ }{\bf##1}
  {\rm (##2) ##3}}}
\def\Journal#1; #2; #3;{\gdef#1##1 (##2) ##3;{{\sl#2\ }%
{\bf##1\kern0.2em#3}\ {\rm (##2) ##3}}}
\def\JOURNAL#1; #2; #3;{\gdef#1##1 (##2) ##3;{{\sl#2\ }%
{\bf#3\kern0.2em##1}\ {\rm (##2) ##3}}}

\journal\Ib; {\it ibid.};

\journal\AA; Astron.\ Astrophys.;
\journal\AAR; Astron.\ Astrophys.\ Rev.;
\journal\AJ; Astron.~J.;
\journal\AJP; Austr.~J.\ Phys.;
\journal\AN; Astr.\ Nachr.;
\journal\ANP; Adv.\ Nucl.\ Phys.;
\journal\ANYAS; Ann.\ N.~Y.\ Acad.\ Sci.;
\journal\ApJ; Astrophys.~J.;
\journal\ApJS; Astrophys.~J.\ Suppl.;
\journal\ApL; Astrophys.\ Lett.;
\journal\APNY; Ann.\ Phys.\ (N.Y.);
\journal\ARAA; Ann.\ Rev.\ Astron.\ Astrophys.;
\journal\ARNS; Ann.\ Rev.\ Nucl.\ Sci.;
\journal\ARNPS; Ann.\ Rev.\ Nucl.\ Part.\ Sci.;
\journal\ASS; Ap.\ Sp.\ Sci.;
\journal\CA; Comm.\ Astrophys.;
\journal\CQG; Class.\ Quantum Grav.;
\journal\CS; Curr.\ Sci.;
\journal\DAN; Dokl.\ Akad.\ Nauk.\ S.S.S.R.;
\journal\FCP; Fund.\ Cosmic Phys.;
\journal\FP; Fortschr.\ Phys.;
\journal\HPA; Helv.\ Phys.\ Acta;
\journal\IJTP; Indian J.\ Theor.\ Phys.;
\journal\JETP; Sov.\ Phys.\ JETP;
\journal\JETPL; JETP Lett.;
\journal\MNRAS; Mon.\ Not.\ R.\ Astron.\ Soc.;
\journal\MSAI; Mem.\ Soc.\ Astron.\ Ital.;
\journal\MPLA; Mod.\ Phys.\ Lett.~A;
\journal\Nature; Nature;
\journal\NC; Nuovo Cim.;
\Journal\NCA; Nuovo Cim.; A;
\Journal\NCC; Nuovo Cim.; C;
\journal\NCS; Nuovo Cim.\ Suppl.;
\journal\NIM; Nucl.\ Instr.\ and Meth.;
\JOURNAL\NPA; Nucl.\ Phys.; A;
\JOURNAL\NPB; Nucl.\ Phys.; B;
\journal\PASJ; Publ.\ Astron.\ Soc.\ Japan;
\journal\PAZ; Pisma Astr.\ Zh.;
\Journal\PLA; Phys.\ Lett.; A;
\Journal\PLB; Phys.\ Lett.; B;
\journal\PPNP; Prog.\ Part.\ Nucl.\ Phys.;
\journal\PTRSLA; Phil.\ Trans.\ Roy.\ Soc.\ London~A;
\journal\PRep; Phys.\ Rep.;
\journal\PR; Phys.\ Rev.;
\journal\PRL; Phys.\ Rev.\ Lett.;
\journal\PRA; Phys.\ Rev.\ A;
\journal\PRB; Phys.\ Rev.\ B;
\journal\PRC; Phys.\ Rev.\ C;
\journal\PRD; Phys.\ Rev.\ D;
\JOURNAL\PRSA; Proc.\ Roy.\ Soc.\ Lond.; A;
\journal\PTP; Prog.\ Theor.\ Phys.;
\journal\PZ; Phys.~Z.;
\journal\PZETF; Pisma Zh.\ Eksp.\ Teor.\ Fiz.;
\journal\RMP; Rev.\ Mod.\ Phys.;
\journal\RNC; Riv.\ Nuovo Cim.;
\journal\RPP; Rep.\ Prog.\ Phys.;
\journal\SA; Sci.\ Am.;
\journal\SAL; Sov.\ Astron.\ Lett.;
\journal\Science; Science;
\journal\SJNP; Sov.\ J.\ Nucl.\ Phys.;
\journal\SP; Solar Phys.;
\journal\SST; Speculations\ Sci.\ Tech.;
\journal\YF; Yad.\ Fiz.;
\journal\ZP; Z.~Phys.;
\journal\ZPC; Z.~Phys.\ C;
\journal\ZETF; Zh.\ Eksp.\ Teor.\ Fiz.;

\nofirstpagenumber
\line{August 1992\hfil MPI-Ph/92-65}
\title{Non-Abelian Boltzmann Equation for Mixing\creturn
and Decoherence\footnote{$^\dagger$}{Based, in part, on work to
be submitted by G.S.\ as a doctoral thesis to the
Ludwig-Maximilians-Universit\"at (Munich).}}

\author{G.~Raffelt, G.~Sigl and L.~Stodolsky}
       {Max-Planck-Institut f\"ur Physik\creturn
       Postfach 401212, 8000 M\"unchen 40, Germany}

\abstract{We consider particle oscillations and their damping in
second-quantized form. We find that the damping or ``decoherence"
may be described by a Boltzmann-like collision integral with
``non-abelian blocking factors" (fermions). Earlier results are
generalized in that the momentum degrees of freedom are included and
that the mixing equations  become intrinsically  non-linear at high
densities.
\medskip
\noindent PACS index categories: 05.20.Dd, 03.65.Bz, 14.60.Gh,
97.60.Bw}

\newpage

\RF\Raffelt{G.~Raffelt, G.~Sigl, and L.~Stodolsky,
    \PRD 45 (1992) 1782;.}
\RF\Harris{R.~A.~Harris and L.~Stodolsky, {\sl J. Chem. Phys.}
    {\bf 74} (1981) 2145 and \PLB 78 (1978) 313;;
    R.~A.~Harris and R.~Silbey, {\sl J. Chem. Phys.} {\bf 78:12}
    (1983) 7330.}
\RF\Thomson{For related attempts to generalize the formalism see
    M.~J.~Thomson, \PRA 45 (1992) 2243;;
    M.~J.~Thomson and B.~H.~J.~McKellar, University of Melbourne
    Preprint UMP 89-108 and Manchester preprint MC-TP-92.}
\RF\krip{For a discussion of this phenomenon in the SU(N) case see
    I.~B.~Khriplovich and V.~V.~Sokolov, {\sl Physica} {\bf A141}
    (1987) 73. A further question one could study by the present
    methods is that of the possiblility of interference effects
    between different degenerate states of large, quasi-macroscopic
    systems; a big molecule, after all, is almost such a system. If
    the system has many equivalent states the SU(N) analysis then
    becomes relevant.}
\RF\Dolgov{A.~Dolgov, \SJNP 33 (1981) 700;, also used such flavor
    densities in the description of neutrino mixing.}
\RF\Abragam{A.~Abragam, {\it The Principles of Nuclear Magnetism}
    (Oxford, 1961), Ch.~viii, gives a standard discussion of spin
    relaxation. Harris and Silbey, Ref.~1, also give numerous
    references on relaxation theory.}
\RF\Stodnu{For neutrinos see L.~Stodolsky \PRD 36 (1987) 2273;,
    an extensive application of the formalism in the early universe is
    given by K.~Enquist, K.~Kainulainen and J.~Maalampi,
    \NPB 349 (1991) 754;.}
\RF\Stodolskya{R.~A.~Harris and L.~Stodolsky, \PLB 116 (1982) 464;.}
\RF\Stodolsky{A review and discussion is given by L. Stodolsky, in:
    J.~S.~Anandan (ed.), {\it Quantum Coherence} (World Scientific,
    Singapore, 1990) pg.~320.}
\RF\pethick{As in N.~Iwamoto and C.~J.~Pethick, \PRD 25 (1982) 313;.}
\RF\Einzel{See for example the discussion of spin effects in
    ${\rm He}^3$ by D.~Einzel, in: M.~J.~R.\ Hoch and R.~H.~Lemmer
    (eds.), {\it Low Temperature Physics},  (Springer, Heidelberg,
    1991) pg.~275.}
\RF\maalampi{J.~Maalampi and J.~T.~Peltoniemi, \PLB 269 (1991) 357;.
    M.~Turner, \PRD 45 (1992) 1066;.}
\RF\raffeltb{G.~Raffelt and G.~Sigl, ``Neutrino Flavor Conversion in a
    Supernova Core", Report MPI-Ph/92-68 (1992).}
\RF\Sawyer{R.~F.~Sawyer, \PRD 42 (1990) 3908;.}

\def\t{\tau\kern-.50em\tau}
\def\Hfree{H_{\rm free}}
\def\Hint{H_{\rm int}}
\def\half{{\textstyle{1\over2}}}
\def\ket#1{\vert#1\rangle}

\def\smean#1{\langle~\vert #1 \vert~\rangle}

\def\G{{\bf G}}
\def\P{{\bf P}}
\def\V{{\bf V}}
\def\p{{\bf p}}
\def\q{{\bf q}}
\def\pp{{{\bf p}'}}
\def\qq{{{\bf q}'}}
\def\x{{\bf x}}
\def\o{\omega}
\def\O{\Omega}
\def\rmed{\rho_{\rm med}}


{\it 1.~Introduction.}---A series of interesting problems, ranging
from the question of ``decoherence", such as for the two states of an
optical isomer\qref{\Harris,\Stodolskya}, spin relaxation in condensed
matter\qref{\Abragam}, or the damping of particle
oscillations\qref{\Stodnu}, pose essentially the same question in a
variety of contexts\qref{\Stodolsky}. For neutrino oscillations in
particular, questions of statistics can  arise, as in the neutrino
degeneracy during the collapsed phase of a supernova. It then becomes
of interest to examine a second-quantized approach to such problems,
where giving the correct commutation relations to the fields takes
care of ``Pauli blocking" and similar  effects. In addition it might
be hoped that the use of field theoretic language may add some insight
into the interesting quantum mechanical issues behind our practical
applications. In attempting such a formulation  we have found a simple
and transparent kinetic equation, which may be thought of as a kind of
``non-abelian Boltzmann equation". Although closely related or partial
versions of this equation seem to exist in various
connections\qref{\Einzel}, we think it useful to present it here in a
general and simple form.

An interesting aspect of the equation  is that it presents kinetic and
``interference" aspects simultaneously. Spin-like problems of the type
under discussion depend very much on subtle phase effects. On the
other hand the Boltzmann type of kinetic equation  seems to be
entirely ``incoherent" and classical in spirit. Nevertheless, we shall
see how a marriage of both aspects arises naturally. This  also may be
of help in understanding many questions including  how the classical
world arises from the quantum mechanical one.

{\it 2.~Matrix of densities.}---In a previous paper\qref\Raffelt, we
gave a field-theoretic approach to mixing problems dealing directly
with the fields. However, incoherence or damping  was not considered.
In the  single-particle framework used previously \qref{\Stodolsky} to
study damping, the object of study (for two-state systems) was the
quantum mechanical $2\times2$ density matrix for the two states. The
two-state system was viewed as imbedded in a much larger system and
the density matrix was found by considering the total wavefunction of
the whole system and tracing over the unobserved variables of the
large system---usually thought of as ``the medium". If the density
matrix was expressed in terms of Pauli matrices $\t$ and a
polarization vector $\P$ as $\half(1+\P\cdot\t)$ then $\P$ was found
to obey a Bloch-like equation: $$\dot\P=\V\times\P-D\P_T.\EQ\bloch$$
This  gives a coherent rotation of the polarization through $\V$, and
through the damping parameter $D$, a loss of coherence or damping
manifested as a shortening of $\P$. The subscript $T$ indicates  those
components non-diagonal (``transverse") with respect to interactions
with the medium.  Also, a formula for $D$ was obtained in $S$-matrix
language, resembling a generalized optical theorem, and giving $D$ in
terms of  the scattering amplitudes on the background. In measurement
theory language, $D$ may be thought of as giving the rate of
``reduction" of the wavefunction of the two-state system
\qref\Stodolsky.

What is to play the role of the quantum mechanical $2\times2$ density
matrix in the field-theoretic formulation? In second-quantized terms
the above procedure of tracing over the unobserved variables may be
accomplished in a sense by applying an annihilation operator to the
state to ``remove" the coordinate  and then finding the overlap with
another such operation. Thus we propose to study  the ``matrix of
densities"
$$\rho_{ij}=\smean{\psi^\dagger_j\psi_i}\EQ\ro$$
instead of the density matrix. The indices refer to the various
components of the field in question and $\ket{~}$ to our original
state. For a two-state system, such as neutrino fields with two
flavors, $\psi$ will be a two-component spinor in flavor space
\qref\Dolgov, and we will have four bilinears to deal with. Diagonal
elements of this c-number matrix are  the ordinary  densities of a
particle type, while the off-diagonal elements contain more subtle
phase information relating to the coherence of the system. The matrix
(\ro) is Hermitian and positive definite. If we assume interactions
that conserve the total number of particles, as we shall in this
paper, then the trace of its spatial integral is conserved.
Hermiticity, positivity and conservation of the trace are the
properties characterizing a density matrix and so it is justified to
speak of the integrated $\rho$ as a density matrix\qref{\Thomson}.

The object (\ro) is of course much more complicated than the density
matrix of the single particle formulation since the arguments of the
fields  can refer to any, and in general different, points of space;
similarly in momentum space they can refer to different momenta.
However, since our major interest is the mixing question, we shall
make the greatly simplifying assumption  of ignoring spatial
inhomogeneities. Our state, $\ket{~}$,  will represent a situation
which is spatially uniform. Thus in position space $\rho$ can only
depend on the difference of the arguments of the fields, while in
momentum space we have
$\smean{\psi^\dagger_j(\pp)\psi_i(\p)}=(2\pi)^3\delta^3(\pp-\p)
\rho_{ij}(\p),$  and so we can confine ourselves to the study of the
matrix $\rho_\p.$ (When matrix indices are suppressed we write $\p$ as
a subscript).

In this paper we shall limit ourselves to  the non-relativistic domain
in the sense that we do not directly consider the role of creation and
annihilation of particles. Thus our fields are not those of fully
relativistic quantum field theory; the operator $\psi(\x)$ contains
only annihilation operators for particles and not the creation
operators for antiparticles.

{\it 3.~Interactions.}---Turning to  the time development of $\rho,$
we treat the fields  as Heisenberg operators $\psi(t)$ and assume that
our constant state $\ket{~}$ is produced by letting the creation
operators for the $\psi$ particles operate on the state vector for the
medium at $t=0$.  This corresponds to a simple product wavefunction
with no correlations between the $\psi$-particles and the medium.  We
follow the approach of  finding the equations among the $\rho_{ij}$ by
turning on the interactions for a short time at $t=0$ and then
assuming these equations to be valid at all times.

Our starting point is the equation of motion for the operator
$\psi^\dagger_j\psi_i$,  the Hamiltonian consisting of a ``free" and
 an interaction term, $H=\Hfree+\Hint$. $\Hfree$ is taken to be
bilinear in the fields and diagonal in momentum. It may describe a
kinetic energy plus some internal interaction of the fields among
themselves. An example is the free Hamiltonian for neutrinos with a
mass matrix, which leads to a rotation in the internal flavor space.
If $(i,j)$ refer to different spin components of a system with a
magnetic moment, $\Hfree$ may refer to an applied magnetic field, also
leading to a rotation among the components. Or for the molecular
isomers $\Hfree$  gives the tunneling between left and right
configurations in addition to the kinetic energy.

$\Hfree$ may be written  for each momentum in terms of a matrix
$\Omega$ and the column vector $\psi$ as
$\psi^\dagger_\p\O_\p\psi_\p$. Commuting with
$\psi^\dagger_j(\p)\psi_i(\p)$  produces  bilinears of the same
momentum. The  set of densities  ``rotate" among each other. This is
of course just  the precession of the ``spin", ordinary or flavor
variety.

$\Hint$ gives the interaction with the background medium which we take
to be  a local  point-like interaction between the $\psi$ and the
background, represented by an operator $B$:
$$\Hint(t)=\int B(x)\psi^\dagger(x)G\psi(x)d^3\x.\EQ\hint$$
$G$ is a coupling constant matrix acting on the column vector of the
$\psi$'s, and $B$ would be typically, as in the Fermi interaction,
bilinear in the background fields. When $G$ is diagonalized it gives a
set of states which scatter only into themselves. The $T$ for
``transverse" in Eq.~(\bloch) then refers to those bilinears which are
off-diagonal in such states. The choice of Eq.~(\hint) is appropriate
to the Fermi interaction and many other problems where the particles
under study undergo a local elastic scattering as their basic
interaction. We shall take $G$ Hermitian, corresponding to lowest
order, but one can imagine relaxing this restriction to represent an
underlying process in higher than Born approximation, as in the
$S$-matrix approach\qref{\Harris,\Stodolskya}.

The commutator of $\Hint$ with $\psi^\dagger_j(\p)\psi_i(\p)$ again
yields a bilinear. Finally taking the expectation value of the
resulting equation of motion yields [notation: $d\p\equiv {d^3\p/
(2\pi)^3}$]:
$$i\dot\rho_{ij}(t,\p)=\left[\O(\p),\rho(t,\p)\right]_{ij}
  +\int d\pp\Bigl\langle\Big\vert B(t,\p-\pp)
  \psi^\dagger_j(t,\p)G_{ik}\psi_k(t,\pp)\Big\vert\Bigr\rangle-{\rm
  h.c.}
  \EQ\rodot$$
This is an exact expression in terms of Heisenberg operators. It
resembles but is not yet in the desired form of a set of equations
among the $\rho_{ij}$, like Eq.~(\bloch), because $B$ cannot be
factorized out of the brackets. The original state $\ket{~}$ is given
by free particle operators, $\psi^\dagger$ at $t=0$, while the
operator expressions in the brackets are  non-trivial in terms of such
Schr\"odinger operators.

We can, however, introduce perturbation theory where we expand the
Heisenberg operators $\psi(t)$ and $B(t)$ in terms of non-interacting
(interaction representation) operators $\psi_0(t)$ and $B_0(t)$
corresponding to $G=0$. A first approximation is to set
$\psi(t)=\psi_0(t)$ and $ B(t)=B_0(t)$. Now with this approximation
the bracket in Eq.~(\rodot) may be factorized since by assumption the
original state $\ket{~}$ has no correlations between the
$\psi$-particles and the background. Only bilinears with equal momenta
survive by virtue of the spatial uniformity represented by $\ket{~}$
and we obtain a closed equation among the $\rho$:
$$i\dot\rho_\p=\left[\O_\p,\rho_\p\right]+
  \rmed\left[G,\rho_\p\right]\,.\EQ\dota$$
Here, $\rmed\equiv\smean{B(0,0)}$ is the medium density, assumed to be
stationary and taken at momentum transfer zero. This relation will be
recognized as adding an ``index of refraction" or medium-induced
energy contribution to the free mixing given by $\Hfree$. Note this
index of refraction effect actually refers to {\it differences} in the
index (as occur for example as in the MSW effect) and does not appear
for a one-component field.

In the two-level [or more generally SU(2)] case where $\O$, $\rho$
and $G$ can be expanded in terms of Pauli matrices [more generally
SU(2) generators], \eg\ $\rho_\p=\half n_\p(1+\P_\p\cdot\t)$, with
coefficients $\V_{\rm free}$, $\P$ and ${\G}$ respectively,
Eq.~(\dota) becomes $\dot\P_\p=\V_\p\times\P_\p$ with
$\V_\p\equiv\V_{\rm free}(\p)+\V_{\rm med}$ where
$\V_{\rm med}=\G\rmed$.
This is the first term of Eq.~(\bloch) and  agrees with the old
result\qref{\Stodolsky}, to first order in $G$.

{\it 4.~Incoherence and damping.}---Now, Eq.~(\dota)  still contains
no sign of any loss of coherence, \ie\ any  shrinking of $\P$. This is
understandable, since to this order of approximation there is no
excitation of the medium. Only forward scattering has been taken into
account, as is indicated by taking $B$ at momentum transfer zero. If
the medium does not change its state, no incoherence can be expected
since the overall wavefunction, assumed to be a simple product at
$t=0$, has remained so. As is familiar in perturbation theory, the
energy but not the wavefunction changes in first order. Thus to see
the first evidence of damping or incoherence we  must go to the next
order.

We stress that spatial homogeneity  excludes sources of damping or
incoherence  due to spatially varying conditions. For example, if the
magnetic field applied to an ensemble of spins varies in space, or if
there are inhomogeneities in the medium in which our neutrinos
propagate \qref{\Sawyer}, there will be an averaging or loss of
coherence in the properties of the ensemble. In general a loss of
coherence can always  result from averaging over some parameters
characterizing the system.  However, here we  only wish to address the
true quantum mechanical loss of coherence.

We go to the next order in $G$ by  expanding the Heisenberg operators
to first order:
$$\eqalign{\psi(t)&=\psi_0(t)+i\int_0^t dt' \left[\Hint(t-
  t'),\psi_0(t)\right],\cr
  B(t)&=B_0(t)+i\int_0^t dt' \left[\Hint(t-
  t'),B_0(t)\right].\cr}\EQ\second$$
To evaluate the new terms in Eq.~(\second) we must now make our
assumptions on the nature of the system explicit. We assume that the
interactions described by $\Hint$ can be taken as individual, isolated
collisions where the particles represented by the $\psi$ go from free
states to free states as in ordinary scattering theory. Thus the time
for a single collision is taken to be short compared to the time
between collisions; furthermore the very small energies represented by
the mixing effects in Eq.~(\dota ) are neglected in calculating the
effects of the collisions. Under these assumptions  we can carry out
what amounts to a Golden Rule calculation on the level of the fields
and find
$$\eqalign{&\psi(t,\p)=\psi_0(t,\p)+ie^{-iE_\p t}
  \int d\pp d\o\delta(E_\pp-E_\p+\o) B_0(\o,\p-
  \pp)G\psi_0(0,\pp),\cr\alignskip &B(t,\p-\pp)=B_0(t,\p-\pp)+
  i\int d\qq d\q d\o\delta(E_\qq-E_\q-\o)\,\times\cr
  &\hskip 4cm\times\,\left[B_0(\o,\qq-\q),B_0(t,\p-\pp)\right]
  \psi^\dagger_0(0,\qq)G\psi_0(0,\q).\cr}\EQ\seconda$$
The $\delta$ function arises from letting the time integrals go to
infinity. We have dropped  an associated principle-part integral
since this has to do with a second order energy shift in the medium,
which we already have to lowest order  in Eq.~(\dota).

We may now substitute Eq.~(\seconda) into the r.h.s.\ of Eq.~(\rodot).
By construction the state $\ket{~}$ has the property that only
bilinear correlations of the type $\smean{\psi^\dagger\psi}$ are
present, so  in carrying out the contractions in the brackets only
$\rho$'s will arise. It must be recognized, however, that an
assumption is involved here, similar to that involved in the classical
Boltzmann equation.  While  we construct the state at $t=0$ such that
no other correlations are present, the philosophy of the method that
the equation found by turning on the interactions for a short time at
$t=0$ can  be used for all times tacitly  assumes that no other kinds
of correlations can build up significantly. In this sense our
derivation is Boltzmann-like and, for example, we exclude the
possibility that our system might have important pair correlations.

Carrying out, then, the contractions with only the survival of
ordinary density bilinears, and evaluating all quantities at $t=0,$
we arrive at our main result:
$$\eqalign{\dot\rho_\p&=-i\left[\O_\p,\rho_\p\right]-
  i\rmed\left[G,\rho_\p\right]\cr\alignskip
  &+\int d\pp\left(S(p'-p){(1-\rho_\p)G\rho_\pp G+{\rm h.c.}\over2}
  -S(p-p'){G(1-\rho_\pp)G\rho_\p+{\rm
  h.c.}\over2}\right)\cr}\EQ\main$$
where $S(\Delta)$, a function of the four-vector $\Delta$, is the
usual dynamical structure function, defined by
$\int_{-\infty}^{+\infty}dt~ e^{i\Delta_0t}
  \smean{B_0(t,{\bf \Delta})B_0(0,-{\bf \Delta})}$.
This is the desired generalization of Eq.~(\bloch) under
Boltzmann-like conditions. It resembles the ordinary Boltzmann
equation with the usual drift term, absent because of spatial
uniformity, replaced by  the commutators for coherent rotations, which
give the free motion in ``spin space".  The principle new features of
Eq.~(\main) relative to Eq.~(\bloch) is the inclusion of the momentum
degrees of freedom and the nonlinearity in $\rho$ introduced by
statistics. The main difference with the ordinary Boltzmann equation
is the matrix structure of the ``collision term". The ``loss" term
contains such effects as the creation of polarization by absorption
while the ``gain" term gives effects like the transfer of polarization
through scattering.

We have taken the original state $\ket{~}$ to be pure. If it is itself
characterized by a density matrix then our various bracketed
expressions can be interpreted as thermal or other averages.
Similarly, if there are several species of background particles,
uncorrelated with one another, their contributions can be summed
separately in the collision integral.  Eq.~(\hint ) is the simplest
possible interaction. However, as long as the  essential feature
that the interactions of the different $\psi$ with the background are
identical up to a coupling constant is retained, we anticipate no
change from more general interactions except for a generalization of
the structure factor $S$. For  example, when the spin structure of
the background is accounted for $S(p-p')$ will be
replaced\qref{\pethick} by the probability $W(p,p')$, which is no
longer simply a function of the difference $(p-p')$. It may be
verified that this has no effect on our arguments.

By integrating over $\p$ we may also study the total density matrix,
with  its  associated total polarization:
$$\dot\rho=-i\rmed\left[G,\rho\right]-
  i\int\left[\O_\p,\rho_\p\right]d\p+ \half\int S(p-
  p')\left[G,\left(\rho_\p G(1-\rho_\pp)-
       {\rm h.c.}\right)\right]d\p d\pp\,.\EQ\tot$$
We see here from the commutator structure that the total number of
particles, given by ${\rm Tr}(\rho)$, is explicitly conserved. Observe
that in the special case where all types scatter with the same
amplitude on the medium, \ie~$G$  proportional to the unit matrix, the
collision term in Eq.~(\tot) vanishes identically. There is then no
dynamical damping, however there remain ``de-phasing" effects  due to
the momentum dependence in the precession  of the different modes.

{\it 5.~Limiting cases and applications:~Ordinary collision
integral.}---The simplest limiting case obtains when we have a
one-component $\psi$ (or equivalently many components but no mixing
terms, or mixing terms which can be simultaneously diagonalized with
$G$) and the background is simply free particles, \ie\
$B(\x)=\phi^\dagger(\x)\phi(\x)$,   $\phi(\x)$ a free field.  Then:
$$S(\Delta)=\int d\q d\qq(2\pi)^4\delta^4(\Delta +q-q')n_{\rm
med}(\q)[1\pm n_{\rm med}(\qq)]\,,\EQ\IIIb$$
where $n_{\rm med}(\q)$ are the occupation numbers in the medium
($\pm$ for bosons or fermions). In the low density limit where the
blocking factor can be neglected, with a  single medium mode $\q_{\rm
med}$, $n(\q)=(2\pi)^3 \rho_{\rm med}\delta^3(\q-\q_{\rm med})$, $S$
becomes simply the  energy $\delta$ function times $2\pi\rho_{\rm
med}$. Because $G^2\int d\p'S$ is the scattering rate,
Eq.~(\main) has the appearance of standard rate-of-gain
and -loss terms. With only a single mode $\p$ of $\psi$ populated, as
for a beam, one finds $\dot\rho_\p/\rho_\p=-G^2\int d\p'S(p-p')$
giving the expected depletion of the beam. $S$ can of course also
account for non-trivial interactions among the background particles;
it will then  have more structure than simply the $\delta$ function.

{\it Bloch equations.}---An opposite, more ``quantum mechanical" limit
occurs when $S$ depends negligibly on the energy transfer, as can
happen, say  for a fixed spin, or more generally when the energy
transfer is small compared to the scale of energies associated with
processes in the medium. Then  $S(\Delta)\approx S({\bf
\Delta})=S({\bf -\Delta})$ and we find that the non-linear terms in
Eq.~(\tot) disappear, leaving for the matrix structure of the
collision term $2G\rho G-GG\rho -\rho GG =-[G,[G,\rho]]$, which puts
the nature of the collision term as a kind of double commutator  in
evidence.  The disappearance  of the non-linear terms is interesting;
self-blocking by the $\psi_i$ does not reduce the damping. Evidently
it should not be thought that the  particles do not scatter even if
they are not permitted to change their state in some particular basis.
It is in this  way, after all, that a degenerate  Fermi gas exerts a
pressure. Blocking effects in the {\it background} as in Eq.~(\IIIb),
on the other hand, are {\it not} to be ignored.

Carrying out the integrations in Eq.~(\tot) gives an equation solely
in terms of the total $\rho$ (for $\O$ independent of $\p$), leading
to a Bloch-like \qref{\Abragam} equation. In the two-level case this
gives for the collision term:
$$\half\int d\p d\p'S({\bf p-p'})n_\p\G \times (\G\times \P)=  -
\half\int d\p d\p' S({\bf p-p'})n_\p\vert\G\vert^2 \P_T=-D\P_T.\EQ\d
$$
$T$ means the direction perpendicular to $\G$, where
$G=\half(G_0+\G\cdot\t)$. This is  the damping term of Eq.~(\bloch).
   The calculations given here verify, by essentially standard
manipulations, the  results of the more intuitive $S$-matrix
arguments, where  $D$ was found for two states (``up" and ``down")
as the real part of $(S_{\rm up}S^\dagger_{\rm down}-1)$ or in terms
of phase shifts to lowest order, $D\sim(\delta_{\rm up}-\delta_{\rm
down})^2$,  which is indeed $\sim(g_{\rm up}-g_{\rm down})^2 $, as in
Eq.~(\d). This contains, for example, the old result
\qref{\Stodolskya} that if one type of particle interacts and the
other not, that the damping parameter is equal to half of the
(unblocked) collision rate of the interacting type.

\newpage

{\it Sign of $D$ and Spontaneous  Polarization.}---$D$ in
Eq.~(\d) has arrived with the correct sign---we have damping and not
anti-damping. This ``arrow of time" arises from straightforward
calculation---at least for the short time the interaction is on. It
is a consequence of our starting from  an uncorrelated state, which
can only become  more correlated in time.
That the damping sign  is not entirely trivial is revealed by
numerical solutions of Eq.~(\main) where we find  when spin and
momentum degrees of freedom are closely coupled  that the relaxation
of the momentum can lead to a transient polarization for an initially
unpolarized situation.

{\it Neutrino Mixing in a Stochastic  Background.}---Flavor mixing in
a supernova would be of significance since it  allows other neutrinos
to share the large $\nu_e$ degeneracy. In previous work the approach
to kinetic and chemical equilibrium was treated in the single-particle
wavefunction approach \qref{\maalampi}. Our kinetic equation, with
degeneracy effects built in, is well suited to this problem.  Even
though our treatment here is not explicitly relativistic, relativistic
kinematics can be incorporated in $\Omega_\p$, and the absence
of the  anti-particle degrees of freedom is not critical since these
are suppressed by the high chemical potential. It is important,
however, to include $\nu_e-e$ conversions; this may be done by
introducing further bilinears involving the electrons, a detailed
account will be given elsewhere \qref{\raffeltb}. For applications
where the anti-particles are important, the formalism should be
extended, which raises the interesting question of dealing with
relativistic fields.

{\it Understanding ``Decoherence".}---We recall \qref{\Stodolsky} that
in the first quantized formulation the value of $\P _T$ gives the
extent to which the eigenstates of $G$, ``up" and ``down" (two-level
case), can interfere. As $\P _T$ is reduced the two states
``decohere". Similarly here, interference terms between operators
$\psi_i$ corresponding to different eigenvalues of $G$  are given by
the appropriate elements of $\rho$  and are generally decreased by the
action of the collision term. Hence we may say that in the collision
integral we are finding the (rate of) ``decoherence". The damping or
decoherence rate is like the scattering rate by the background,
but with the difference amplitude $\sim(g_{\rm up}-g_{\rm down})$;
{\it difference} because  the scattering must distinguish the two
systems in order to ``measure" and ``reduce" them.  Hence interactions
not distinguishing the states in question are not ``reducing". The
other extreme, that of strong damping, leads through the rapid
disappearance of $\P_T$, to the ``freezing" of the other
component\qref{\krip} and hence to the ``Turing-Xeno-Watched-Pot"
effect\qref{\Stodolskya}, giving the stabilization\qref{\Harris} of
the molecular isomers.

We are grateful to W.~G\"otze for extensive conversations on
relaxation theory, to D.~Einzel for a discussion concerning spin
effects in ${\rm He}^3$ and to P.~Breitenlohner for a helpful remark.

\refout

\bye